\documentclass[twocolumn,aps,prl,10pt,amsmath,amssymb,nofootinbib,showpacs,superscriptaddress,floatfix]{revtex4-1}

\DeclareFontFamily{U}{rcjhbltx}{}
\DeclareFontShape{U}{rcjhbltx}{m}{n}{<->rcjhbltx}{}
\DeclareSymbolFont{hebrewletters}{U}{rcjhbltx}{m}{n}

\usepackage{graphicx}
\usepackage{color}
\usepackage[usenames,dvipsnames]{xcolor}
\usepackage[colorlinks=true,linkcolor=Red,citecolor=Green,linktoc=page]{hyperref}
\usepackage{multirow}
\usepackage{float}
\usepackage{flushend}
\usepackage{balance}
\usepackage[varg]{txfonts}
\usepackage{ulem}
\usepackage{fancyhdr}

\DeclareMathSymbol{\lamed}{\mathord}{hebrewletters}{108}

\begin{document}
\title{Gauge Theories of Josephson Junction Arrays: Why Disorder Is Irrelevant for the Electric Response of Disordered Superconducting Films}

\begin{abstract}
We review the topological gauge theory of Josephson junction arrays and thin film superconductors, stressing the role of the usually forgotten quantum phase slips, and we derive their quantum phase structure. A quantum phase transition from a superconducting to the dual, superinsulating phase with infinite resistance (even at finite temperatures) is either direct or goes through an intermediate bosonic topological insulator phase, which is typically also called Bose metal. We show how, contrary to a widely held opinion, disorder is not relevant for the electric response in these quantum phases because excitations in the spectrum are either symmetry-protected or neutral due to confinement. The quantum phase transitions are driven only by the electric interaction growing ever stronger. First, this prevents Bose condensation, upon which out-of-condensate charges and vortices form a topological quantum state owing to mutual statistics interactions. Then, at even stronger couplings, an electric flux tube dual to Abrikosov vortices induces a linearly confining potential between charges, giving rise to superinsulation.
\end{abstract}



\author{C.\,A.\,Trugenberger}

\affiliation{SwissScientific Technologies SA, rue du Rhone 59, CH-1204 Geneva, Switzerland}


\
\begin{abstract}
	\noindent

\end{abstract}
\maketitle


\section{Josephson Junction Arrays as a Model for Planar Superconductors}

Superconductivity (for a review, see \cite{tinkham}) is predicated on the formation of a ground state in which magnetic interactions are much stronger than electric ones. When the thickness of a superconducting film is decreased, however, electric interactions become stronger and stronger, until the superconducting ground state is destroyed in favour of different phases \cite{efetov, haviland, fisher}. These phase transitions, which can also be driven by an applied magnetic field at fixed (small) thickness, are accompanied by the dissociation of the material into an emergent granular structure of islands of charge condensate \cite{sacepe}. In this configuration, superconductivity is due to Josephson tunnelling between islands when global phase coherence is established. The other phases form when global phase coherence is lost. 

These granular superconductors near their quantum phase transition can be modelled by Josephson junction arrays (JJAs) \cite{fazio}, fabricated regular quadratic lattices of superconducting islands deposited on a substrate and coupled by Josephson junctions (for a review, see \cite{zant}). The phase transitions are typically driven by tuning the Josephson coupling $E_J$ with respect to a mostly fixed charging energy $E_C$ of the islands. In some of the most recent implementations, e.g., the Al islands are deposited on a semiconductor InAs substrate and the Josephson coupling is driven by a voltage gate \cite{marcus}. Typical island dimensions are $O\left( 1 \mu m\right)$, while their typical distances are $O\left( 100 \ nm\right)$.

It is often believed that the quantum transition destroying superconductivity is due to the disorder embodied by the irregular granularity \cite{feigelman}. Here, we show that it is not so; this quantum transition is caused by electric interactions becoming ever stronger, and this increase in the electric coupling has nothing to do with disorder, as can be witnessed in the fact that it also occurs in perfectly regular JJAs.  

{In discussions of disorder, one starts from a disorder-free Hamiltonian, which defines the spectrum of excitations. Then, one lets some parameters of this Hamiltonian become random variables, be it the potential energy of electrons, as in the original formulation \cite{anderson}, the Ising couplings, when discussing spin glasses (for a review, see \cite{parisi}) or other couplings in general. For planar superconductors, the appropriate disorder-free Hamiltonian to start with is that of Josephson junction arrays, which, in their classical limit, are embodiments of the XY model, which is the paradigm of 2D superconductivity (for a review, see \cite{minnhagen}). The disorder to be added consists then in allowing random sizes (and shapes) of the local condensate islands and random distances between them (the graph connectivity is typically held fixed in these discussions to maintain the existence of the superconducting phase). This disorder models the typical structure of irregular islands of condensate characterizing superconducting films \cite{sacepe}, and is reflected in the couplings $E_J$ and $E_C$ becoming random variables centred around their typical JJA values. }

Starting from the correct disorder-free Hamiltonian ensures that one deals with the correct spectrum of excitations. It is these excitations that are affected by the disorder embodied in the random couplings. For example, in the 1D disordered Schwinger model, one should start from the bosonized version of the Hamiltonian, since linear confinement implies that the spectrum consists only of neutral mesons \cite{sondhi}. In the present case, the correct disorder-free system precisely consists of JJAs. As we now show, the phases obtained in this system when superconductivity is destroyed are either topological, and hence disorder becomes ``transparent'', or do not have charged excitations in the spectrum, and disorder can thus influence possibly thermal properties but not electric transport. This has the consequence that they are genuine new phases of matter. 

In a nutshell, what happens is as follows. On a 2D JJA, quantum phase slips are equivalent to vortex tunnelling on the dual array. When vortices can tunnel, charges and vortices can both be out of condensate simultaneously. But as soon as this happens, charges and vortices are frozen into a topological ground state by their mutual statistics interactions: this is the origin of the Bose metal. Moreover, when vortex tunnelling events proliferate, electric fields are squeezed into electric flux tubes between charges and holes. This causes a linearly confining potential and the infinite resistance of the superinsulating phase. 

However, there is one important difference between superconducting films and JJAs. Films are made of one single material, it is only the condensate that dissociates into superconducting islands; JJAs typically have physical islands of a different material than the exposed substrate between them. As we will show, this implies that to see all phases in JJAs, one most probably needs two driving parameters, with one setting the charge tunnelling strength $E_J$ between the islands and one governing the vortex tunnelling strength $E_C$ on the ``dual array'' between them. 

\section{Josephson Junction Arrays: The Standard Treatment}

Josephson junction arrays (for a review, see \cite{zant}) are quadratic arrays of spacing $\ell$ of superconducting islands with nearest neighbours Josephson couplings of strength $E_J$. Each island has a capacitance $C_0$ to the ground and a mutual capacitance $C$ to its neighbours. The Hamiltonian for such a system is
\begin{equation}
H=\sum_{\bf x} \ {C_0\over 2} V_{\bf x}^2 + \sum _{<{\bf x \bf y}>}
\left( {C\over 2} \left( V_{\bf y}-V_{\bf x} \right) ^2 + E_J
\left( 1-{\rm cos}\ \left( \varphi _{\bf y} - \varphi _{\bf x} \right) \right)
\right) \ ,
\label{ham1}
\end{equation}
where boldface characters denote the sites of the array, $<{\bf x \bf y}>$ indicates nearest neighbours, $V_{\bf x}$ is the electric
potential of the island at ${\bf x}$, and $\varphi _{\bf x}$ the phase of its order parameter.  Introducing the forward and backwards lattice derivatives (which are exchanged under summation by parts): 
\begin{eqnarray}
\Delta_i f({\bf x}) = f({\bf x} + \hat i) -f({\bf x}) \ ,
\nonumber \\
\hat \Delta_i f({\bf x}) = f({\bf x})-f({\bf x} - \hat i) \ ,
\label{latder}
\end{eqnarray}
where $\hat i$ denotes a unit vector in direction $i$, and the corresponding lattice Laplacian  $\nabla^2 = \hat \Delta_i \Delta_i$, we can 
rewrite the Hamiltonian as
\begin{equation}
H= \sum_{\bf x} \ {1\over 2} V_{\bf x} \left( C_0 - C\nabla^2 \right) V_{\bf x} +
\sum _{{\bf x},i} \ E_J \left( 1-{\rm cos} \  \left( \Delta_i \varphi_{\bf x} \right)
\right) \ ,
\label{ham2}
\end{equation}
where we have chosen natural units so that $c=1$, $\hbar = 1$, $\varepsilon_0=1$ and we have set the lattice spacing $\ell=1$ for ease of presentation. 

The phases $\varphi _{\bf x}$ are quantum-mechanically conjugated to the charges (Cooper pairs) 
$2e q_{\bf x}\ ,\, q_{\bf x} \in Z $ on the islands, where $e$ is the electron charge. The Hamiltonian (\ref{ham2}) can be expressed in terms of charges and phases by noting that the electric potentials $V_{\bf x}$ are determined by the charges $2e q_{\bf x}$ via a discrete version of Poisson's equation:
\begin{equation}
\left( C_0 -C\nabla^2 \right) V_{\bf x} = 2e q_{\bf x} \ .
\label{poisson}
\end{equation}

Using this in (\ref{ham2}) we obtain
\begin{equation}
H= \sum_{\bf x} \ 4 E_{\rm C}^{\prime}\ q_{\bf x} {1\over {C_0/C}-\nabla^2 } q_{\bf x} +
\sum _{{\bf x},i} \ E_{\rm J} \left( 1-{\rm cos} \ \left( \Delta _i \varphi_{\bf x} \right)
\right) \ ,
\label{hamfin}
\end{equation} 
where $E_{\rm C}^{\prime}\equiv e^2/2C$. The integer charges $q_{\bf x}$ interact via a two-dimensional Yukawa potential of mass $\sqrt{C_0/C} $. We have denoted the coupling of this potential as $E_{\rm C}^{\prime}$ to signify that, contrary to what is normally assumed, it is not the full charging energy $E_{\rm C}$ of the islands, as we will now show. 

The partition function of the JJA admits a phase-space path-integral representation \cite{fazio} 
\begin{eqnarray}
Z &&= \sum_{\{q_{\bf x}\}} \int_{-\pi }^{+\pi } {\cal D}\varphi
\ {\rm exp}(-S)\ ,
\nonumber \\
S &&=\int_0^{\beta}dt  \sum _{\bf x}  i\ q_{\bf x} \ \dot \varphi_{\bf x} + 4 E_{\rm C}^{\prime} \ q_{\bf x} {1\over {C_0/C}-\nabla^2 } q_{\bf x}
\nonumber \\
&&+\sum _{{\bf x}, i} E_J \left( 1- {\rm cos}\ \left( \Delta _i \varphi_{\bf x} \right)
\right)\ ,
\label{partfunc}
\end{eqnarray} 
where $\beta = 1/T$ is the inverse temperature. In (\ref{partfunc}), (Euclidean) time has to be considered also as discrete, as generally appropriate when degrees of freedom can change only in integer steps. We introduce thus a discrete time step $\ell_0$, which is the typical time scale associated with tunnelling events. We thus substitute the time integrals and space sums over a lattice with nodes ${\bf x}$ by a sum over space-time lattice nodes $x$, with $x^0=t$ denoting the discrete time direction. Also, in what follows we shall consider the purely quantum theory at zero temperature by letting $\beta \to \infty$; incorporating a finite temperature is easily performed by restricting the time sums to a finite domain. Denoting the (forward) finite-time differences by $\Delta_0$, we obtain 
\begin{eqnarray}
Z &&=\sum_{\{j_0\}} \int_{-\pi }^{+\pi } {\cal D}\varphi
\ {\rm exp}(-S)\ ,
\nonumber  \\
S&&=\sum _x  i\ j_0 \Delta_0 \varphi + 4 \ell_0 E_{\rm C}^{\prime} \ j_0 {1\over {C_0/C}-\nabla^2 } j_0
\nonumber \\
&&+\sum _{x, i} \ell_0 E_{\rm J} \left( 1- {\rm cos}\ \left( \Delta _i \varphi \right) \right) \ ,
\label{part}
\end{eqnarray} 
where we now denote the integer charge degrees of freedom by $j_0$ for reasons to become clear in a moment. Going over to the Villain representation, the partition function can be formulated as
\begin{eqnarray}
Z &&=\sum_{\{ a_i\}, \{j_0\}} \int {\cal D} j_i \int_{-\pi }^{+\pi } {\cal D}\varphi
\ {\rm exp}(-S)\ ,
\nonumber  \\
S&&=\sum _{x,i}  i\ j_0 \Delta_0 \varphi + ij_i \left( \Delta_i \varphi + 2\pi a_i \right) + 
4 \ell_0 E_{\rm C}^{\prime} \ j_0 {1\over {C_0/C}-\nabla^2 } j_0
+{1\over 2\ell_0 E{\rm J}} j_i^2  \ ,
\label{part}
\end{eqnarray} 
where $a_i \in {\mathbb Z}$ are integers and $j_i$ represents the total charge current in direction $\hat i$. 

\section{Adding Quantum Phase Slips}
In 1D Josephson junction chains, quantum phase slips \cite{golubev} are crucial in the regime of low temperatures and, accordingly, they are routinely taken into account (for a review, see \cite{arutyunov}). For some unknown reason, however, the corresponding tunnelling events are mostly neglected for 2D JJAs, which leads to the wrong results. Let us first note that $a_i $ in (\ref{part}) constitutes a lattice gauge field. If we make a transformation $a_i \to a_i +\Delta_i \lambda$, with an integer $\lambda$, we can absorb $\lambda$ into $\varphi$ with a shift by a multiple of $2\pi$ of its integration domain. We can then shift the definition of the integers $a_i$ to re-establish the original integration domain and the original action, showing that the gauge transformation does indeed leave the partition function invariant. As a consequence, only the transverse, pseudo-scalar components of $a_i$ constitute gauge-invariant quantities; these are the vortices in the model. Note two very important, and also often overlooked, facts: these are core-less vortices, characterized only by their gauge structure, i.e., the circulation of the phases around an array plaquette. As such they can tunnel without dissipation from one site of the dual array to a neighbouring one. Moreover, contrary to Cooper pairs, they are not Noether charges but purely topological ones. Since they have no core, they can not only tunnel from one site of the dual array to another, but they can also appear/disappear on one site in tunnelling events that change the topological quantum number, called instantons (for a review see \cite{coleman}). 

\begin{figure}[H]
\includegraphics[width=8.5 cm]{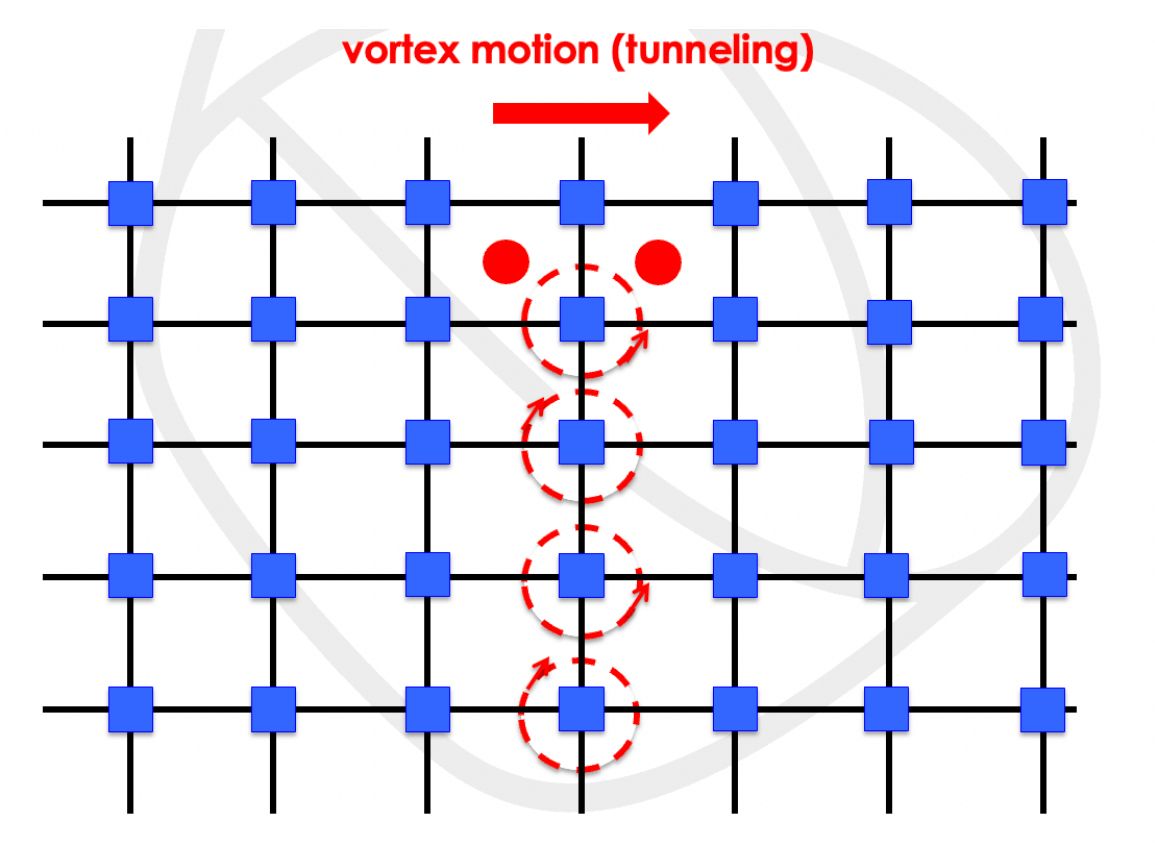}
\caption{A half-line of simultaneous and alternating $2\pi$ phase flips. The only gauge-invariant degree of freedom is the quantum phase slip at the endpoint, corresponding to the displacement of a vortex from one plaquette based on this endpoint to the adjacent one. \label{fig1}}
\end{figure}

In 1D Josephson junction chains, quantum phase slips are local quantum tunnelling events in which the phase of the condensate at one particular island undergoes a $2\pi$ flip over the typical time scale $\ell_0$. In 2D JJAs, instead, quantum phase slips are half-lines of simultaneous such flips of alternating chirality, ending in one particular island, as shown in Figure~\ref{fig1}. These configurations are nothing other than half-lines in which the gauge field $a_i$ alternatingly increases or decreases by one unit. Since there are nowhere vortices but on the two plaquettes based on the endpoint, the only gauge-invariant quantity in this configuration is the quantum phase slip at the endpoint, corresponding to the displacement of a vortex from one plaquette based there to the adjacent one, as shown in Figure~\ref{fig1}. Thus, again, we have gauge-invariant instantons. These must be taken into account at low temperatures. 

Quantum phase slips in JJAs amount thus to vortex tunnelling on the dual array (centres of the plaquettes), the dual phenomenon to Cooper pair tunnelling on the array. This tunnelling {\it cannot be neglected}
 when studying the quantum phase structure of the arrays. We must thus add to the action a vortex kinetic term dual to the corresponding tunnelling current for Cooper pairs \cite{dst, jja}. Of course, this has an important consequence: a vortex moving in one direction creates a 2D logarithmic Coulomb potential between the islands in the perpendicular direction. 

Before showing how this can be incorporated into the action we have to pause a moment to introduce the gauge-invariant lattice version of the curl operator  $\epsilon_{\mu \alpha \nu} \partial_{\alpha}$, where the Greek letters denote the three possible directions on the Euclidean 3D lattice, with sites denoted by $\{ x \}$ \cite{ref-dst}. We first introduce forward and backward finite differences also in the (Euclidean) time direction,
\begin{eqnarray}
\Delta_0 f(x) &&= { f(x +\ell_0 \hat \mu) -f(x) \over \ell_0} \ , \nonumber \\
\hat \Delta_0 f(x) &&= { f(x) -f(x -\ell_0\hat \mu) \over \ell_0} \ .
\label{fttime}
\end{eqnarray}
Then, we introduce forward and backwards shift operators  
\begin{eqnarray}
S_{\mu } f(x) &&= f(x +d \hat \mu) \ ,
\nonumber \\
\hat S_{\mu } f(x) &&= f(x -d \hat \mu) \ ,
\label{shift}
\end{eqnarray}
where $\hat \mu$ denotes a unit vector in direction $\mu$ and $d=1$ in the spatial directions, $d=\ell_0$ in the Euclidean time direction. Summation by parts on the lattice interchanges both the two finite differences (with a minus sign) and the two shift operators. Gauge transformations are defined by using the forward finite differences. In terms of these operators, one can then define two lattice curl operators
\begin{equation}
K_{\mu \nu} = S_{\mu} \epsilon_{\mu \alpha \nu} \Delta_{\alpha} \ , \qquad \qquad \hat K_{\mu \nu} = \epsilon_{\mu \alpha \nu} \hat \Delta_{\alpha} \hat S_{\nu} \ ,
\label{latticecs}
\end{equation}
where no summation is implied over the equal indices $\mu$ and $\nu$. 
Summation by parts on the lattice interchanges also these two operators (without any minus sign). Gauge invariance is then guaranteed by the relations
\begin{equation}
K_{\mu \nu} \Delta_{\nu} = \hat \Delta_{\mu} K_{\mu \nu} = 0 \ , \qquad \qquad \hat K_{\mu \nu} \Delta_{\nu} = \hat \Delta_{\mu} \hat K_{\mu \nu} = 0 \ .
\label{gaugeinv}
\end{equation}
Note that the product of the two curl operators gives the lattice Maxwell operator
\begin{equation}
K_{\mu \alpha} \hat K_{\alpha \nu} = \hat K_{\mu \alpha} K_{\alpha \nu} = -\delta_{\mu \nu} \Delta +\Delta_{\mu} \hat \Delta_{\nu} \ ,
\label{maxwell}
\end{equation}
where $\Delta = \hat \Delta_{\mu} \Delta_{\mu}$ is the 3D Laplace operator. 

Using this notation, the vortex number density is $\phi_0= K_{0i} a_i$. Since the vortex three-current is conserved, we can write it completely in terms of a gauge field $a_{\mu}$ after introducing a Lagrange multiplier $a_0$: $\phi_{\mu} = K_{\mu \nu} a_{\nu}$. The partition function, including quantum phase slips, is then
\begin{eqnarray}
Z&&=\sum_{\{ a_\mu \}, \{ j_0 \} }   \int {\cal D} a_0  {\cal D} j_i  \int_{-\pi }^{+\pi } {\cal D}\varphi
\ {\rm exp}(-S)\ ,
\nonumber \\
S&&=\sum _x  i\ j_0 \left( \Delta_0 \varphi +2\pi a_0\right)  + i j_i \left( \Delta_i \varphi + 2\pi a_i \right) 
+ {1\over 2\ell_0 E_{\rm J}} {j_i}^2 + {\pi^2 \over 4\ell_0 E_{\rm C}} \phi_i^2 \ .
\label{partition2}
\end{eqnarray} 
The Gauss law associated with the Lagrange multiplier $a_0$ leads, in the Coulomb gauge, to the 2D Coulomb interaction term 
\begin{equation}
S_{\rm Coulomb} = \sum_{x} \dots + 4\ell_0 E_{\rm C} j_0 {1\over -\nabla^2} j_0 +\dots  \ ,
\label{cou}
\end{equation}
for charges. This shows that, on sufficiently large samples, the charging energy of the islands is dominated by the 2D logarithmic Coulomb interaction associated with vortex tunnelling (quantum phase slips) and not by the screened interaction due to the finite capacitances. This can lead only to finite-size corrections to the dominant $E_C$ in (\ref{partition2}). 

At this point, we note that the charge current $j_{\mu}$ is conserved and, hence, it can be represented as the dual field strength associated with a second emergent gauge field $b_{\mu}$ as $j_0 = K_{0i}b_i$, $j_i= K_{i0}b_0 + K_{ij}b_j$, where $b_0$ is a real variable, while $b_i$ are integers. We then use Poisson's formula,
\begin{equation}
\sum_{n_{\mu}} f\left( n_{\mu} \right) = \sum_{k_{\mu}} \int dn_{\mu} f\left( n_{\mu} \right) {\rm e}^{i2\pi n_{\mu}k_{\mu}} \ ,
\label{poisson}
\end{equation}
turning a sum over integers $\{ n_{\mu} \}$ into an integral over real variables, to make all components of the gauge fields $a_{\mu}$ and $b_{\mu}$ real, at the price of introducing integer link variables $Q_i$ and $M_i$,
\begin{eqnarray}
Z &&= \sum_{\{ Q_i \} }\sum_{\{ M_i \} }  \int {\cal D} a_{\mu} {\cal D} b_{\mu} \int_{-\pi }^{+\pi } {\cal D}\varphi \ {\rm exp}(-S)\ ,
\nonumber \\
S&&=\sum _{x}  i 2\pi \ a_{\mu} K_{\mu \nu} b_{\nu}+{1\over 2\ell_0 E_{\rm J}} {j_i}^2 + {\pi^2 \over 4\ell_0 E_{\rm C}} \phi_i^2 + i 2\pi a_i Q_i + i 2\pi b_i M_i 
\nonumber \\
&&+ b_i \left( \hat K_{i0} \Delta_0 \varphi + \hat K_{ij} \Delta_j\varphi \right) 
+b_0 \hat K_{0i}\Delta_i \varphi \ .
\label{partition3}
\end{eqnarray}

Finally, we note that the quantities $\hat K_{\mu \nu} \Delta_{\nu} \varphi $ are the circulations of the array phases around the plaquettes orthogonal to the direction $\mu$ in 3D Euclidean space-time, and are thus quantized as $2\pi \ {\rm integers}$. We can thus absorb the quantities $\left( \hat K_{i0} \Delta_0 \varphi + \hat K_{ij} \Delta_j\varphi \right)$ in a redefinition of the integers $M_i$, and define $\hat K_{0i} \Delta_i \varphi = 2\pi M_0$. The original integral over the phases $\varphi$ can then be traded for a sum over the vortex numbers $M_0$, 
\begin{eqnarray}
Z &&= \sum_{\{ Q_i \} }\sum_{\{ M_{\mu}\} }  \int {\cal D} a_{\mu} {\cal D} b_{\mu} \ {\rm exp}(-S)\ ,
\nonumber \\
S &&=\sum _{x}  i 2\pi \ a_{\mu} K_{\mu \nu} b_{\nu}+{1\over 2\ell_0 E_{\rm J}} {j_i}^2 + {\pi^2 \over 4\ell_0 E_{\rm C}} \phi_i^2 + i 2\pi a_i Q_i + i 2\pi b_{\mu} M_{\mu} \ .
\nonumber \\
\label{partition3}
\end{eqnarray} 

This is the gauge theory of JJAs \cite{dst, jja}. The quantities $Q_i$ and $M_i$ represent the Josephson currents of Cooper pairs and the dual vortex tunnelling currents, respectively, while $M_0$ is the vortex number. Charges and vortices interact via two emergent gauge fields $a_{\mu}$ and $b_{\mu}$. The first, infrared-dominant term in the gauge action is the lattice version of the topological Chern--Simons term \cite{jackiw}. The remaining two terms are quadratic in the emergent electric fields, which are orthogonal to the total charge and vortex currents. These are the Josephson tunnelling currents plus local fluctuation terms deriving from time-dependent vortex currents for charges, and vice versa for vortices. The action is thus a non-relativistic version of the Maxwell--Chern--Simons gauge action corresponding to infinite magnetic permeability, in which emergent magnetic fields are suppressed. Gauge fields are topologically massive \cite{jackiw}, the topological gap coinciding with the Josephson plasma frequency of the array, $\omega_{\rm P} = \sqrt{8E_JE_C}$. The dimensionless parameter $g=\sqrt{\pi^2 E_J/2E_C}$, measuring the relative strength of magnetic and electric interactions, drives the quantum phase structure. Note that the quantity $E_C$ here is different from the previously introduced $E_C^{\prime}$: the former is the long-range logarithmic potential induced by vortex tunnelling, the latter is the sub-dominant short-range component arising from the island capacitances. 

\begin{figure}[H]
\includegraphics[width=8.5 cm]{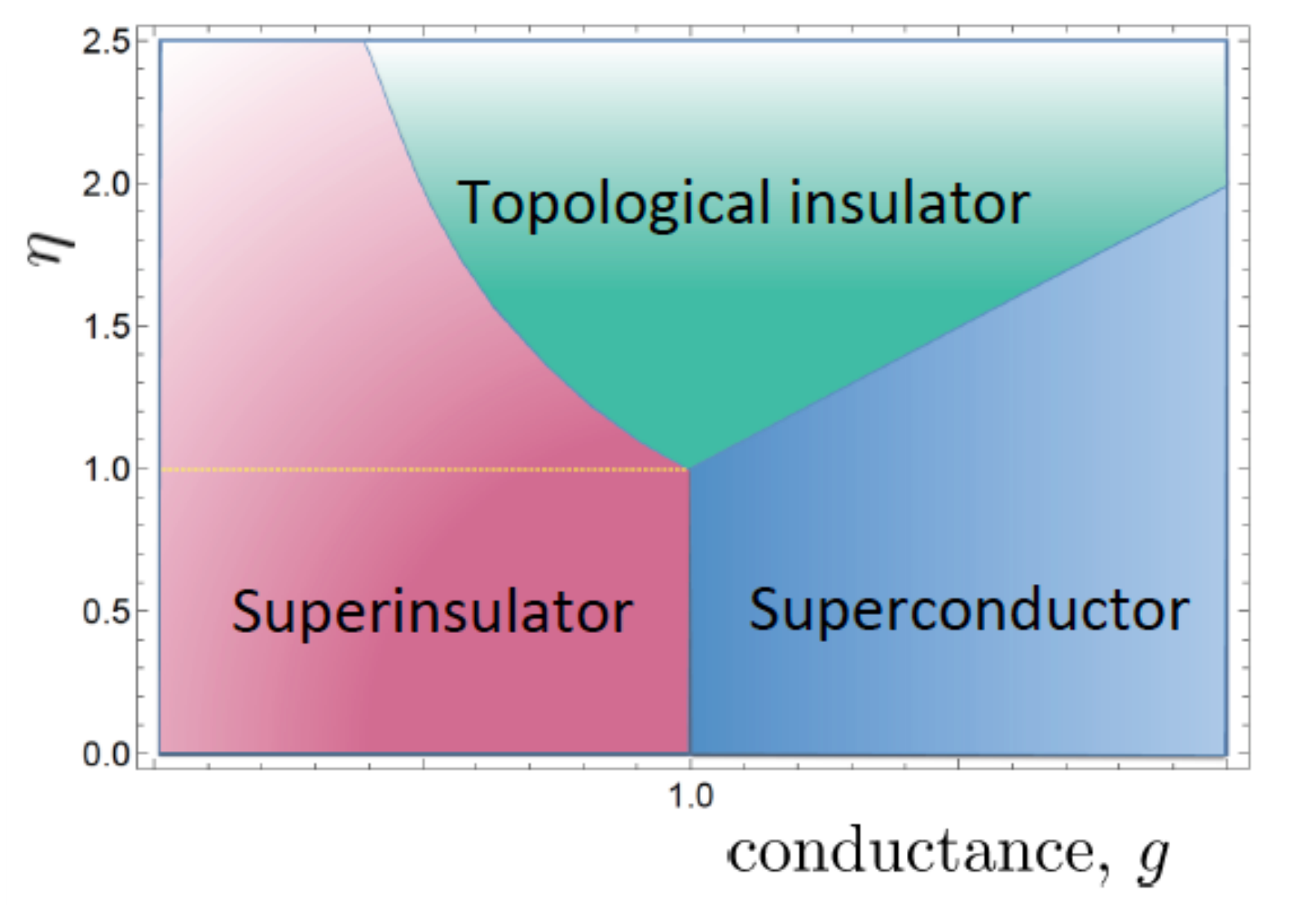}
\caption{The quantum phase structure of JJAs. \label{fig2}}
\end{figure}

The quantum phase structure is determined by the condensation, or lack thereof, of integer electric ($Q_i$) or magnetic ($M_{\mu}$) strings on a 3D Euclidean lattice \cite{dst}. Both their energy and their entropy are proportional to their length and the condensations are governed by energy/entropy balance conditions. The resulting quantum phase structure is shown in Figure~\ref{fig2}.

In addition to the dimensionless conductance parameter $g$, there is another relevant parameter $\eta$ \cite{dst}, which is a function of the ratio $\omega_{\rm P} \ell_0$ of the two characteristic frequencies in the problem, the plasma frequency and the tunnelling frequency $1/\ell_0$. When $\eta <1$, there is a direct ``first-order'' (coexistence of two phases) quantum transition between superconducting and superinsulating phases as $g$ is decreased below the resistance quantum at $R_Q= 6.45 \ k\Omega$ at $g=1$, i.e., electric interactions become stronger. When $\eta > 1$, an intermediate Bose metal state \cite{dst} (for a review, see \cite{phillips, kapitulnik1})
appears in the interval $1/\eta < g < \eta$; as we now review \cite{bm}, this phase is actually a bosonic topological \mbox{insulator \cite{lu, senthil}.} Exactly this quantum phase structure has been recently derived experimentally in an irregular In/InO composite array, although the driving magnetic field was too low to detect superinsulation \cite{kapitulnik2}.

When magnetic interactions dominate ($E_J \gg E_C$), electric strings $Q_i$ condense. This means that Josephson currents of charges percolate through the sample, establishing global superconductivity. This superconductivity, with dissipationless vortices whose confinement/deconfinement drives the thermal phase transition to a resistive state is a type-III superconductivity, not described by the usual Ginzburg--Landau theory \cite{planar}, and is not confined to 2D but exists also in 3D \cite{typeiii}. It has also been proposed to described the physics of high-$T_c$ cuprates \cite{dyons}. We shall not discuss this phase further here, but we shall rather focus on what happens when superconductivity is destroyed by strong electric interactions. 

For $\eta >1$, there is an intermediate domain in which neither electric nor magnetic strings condense: no superconducting currents, and vortices are gapped excitations. In this intermediate domain, the infrared-dominant action for the the JJA reduces to the topological Chern-Simons term
\begin{equation}
S_{\rm TI} =\sum _{x}  i 2\pi \ a_{\mu} K_{\mu \nu} b_{\nu} \ .
\label{topins}
\end{equation}

This is the action of a bosonic topological insulator \cite{lu, senthil}. In this phase, both charges and vortices are frozen in the bulk. The ground state wave function of this state has been derived in \cite{nodisorder}; it consists of an integer-filling composite quantum incompressible fluid of charges and vortices at $g=1$ with excess charges and vortices for $g\ne 1$ forming a Wigner crystal, with charges being in excess of vortices for $g>1$ and vice versa. 

While the bulk is completely frozen at zero temperature, there remain edge currents, where edges may be internal to the sample, forming a percolation structure \cite{cc}. On these edges, we also have the usual 1D quantum phase slips \cite{golubev, arutyunov}, corresponding to vortices moving across the edges. These cause the observed metallic saturation of the resistance, which is the origin of the name ``Bose metal'' for this phase, first predicted in \cite{dst}. Of course, since there is an overabundance of charges for $g>1$, the resistance is lower here than in the region $g<1$ where we have an overabundance of vortices. Correspondingly, when the temperature is raised, the resistance increases in the regime $g>1$, which is the origin of the alternative name ``failed superconductor'', while it decreases in the opposite regime $g>1$, giving rise to the alternative name ``failed insulator''. Failed superconductors and failed insulators, however, are two faces of the same medal, the intermediate bosonic topological insulator \cite{bm}. 

The bosonic topological insulator is the physical embodiment of a field-theoretic anomaly involving Chern--Simons gauge fields \cite{djt}. In topologically massive gauge theories, the limit $m\to \infty$ does not commute with quantization because of the phase space reduction this limit entails \cite{djt}. In physical applications, the topological gauge theory in (\ref{topins}) must always be considered as the $m \to \infty$ limit of the full theory (\ref{partition3}) with dynamical terms; otherwise, wave functionals would not be normalizable. This implies, in particular, that phase and charge are {\it not} a canonically conjugate pair, as would follow from the pure Chern--Simons term (\ref{topins}). Therefore, charges and vortices can be both out of condensate, even when their gap is finite and they can be excited in the bulk. Of course, at $T=0$, they are immediately frozen into a topological ground state by the mutual statistics interactions, giving rise to the topological insulator/Bose metal phase via edge transport \cite{bm}. 

When electric interactions dominate ($E_C \gg E_J$), there is a condensation of vortices while Josephson currents are suppressed. To establish the nature of this phase, let us couple the total electric current $j_{\mu}$ to the real electromagnetic gauge field $A_{\mu}$,
\begin{equation}
S \to S+ i \sum_x A_{\mu} j_{\mu} = S+ i\sum_x A_{\mu} K_{\mu \nu} b_{\nu} \ ,
\label{resp1}
\end{equation}
and integrate over the emergent gauge fields $a_{\mu}$ and $b_{\mu}$ to obtain the electromagnetic effective action. In the 
limit $\ell_0 \omega_P \gg 1$, this is given by
\begin{equation}
S_{\rm eff} \left (A_{\mu}, M_i \right) =
{g \over 4\pi \ell_0 \omega_P }  \ \sum_x \left( F_i-2\pi M_i \right)^2   \ ,
\label{resp2}
\end{equation}
where $F_i$ are the spatial components of the dual electromagnetic vector strength $F_{\mu } = \hat K_{\mu \nu} A_{\nu}$, and where the integers $M_i$ have to be summed over in the partition function. This is a deep non-relativistic version of Polyakov's compact QED action \cite{polyakov, polyakovbook}, in which only electric fields survive. Its form shows that the action is periodic under shifts $F_i \to F_i + 2\pi N_i$, with integer $N_i$, and that the gauge fields are thus indeed compact, i.e., angular variables defined on the interval $[-\pi, +\pi]$. The integers $M_i$ can be decomposed into transverse and longitudinal components, of which neither alone has to be an integer, only the sum is so constrained. The transverse components can be absorbed into a redefinition of $F_i$; the longitudinal components can be represented as
\begin{equation}
M_i^{\rm L} = {\Delta_i \over \nabla^2} m \ ,
\label{mono}
\end{equation}
where $m \in {\mathbb Z}$ are magnetic monopole instantons \cite{polyakov, polyakovbook}. These represent tunnelling events in which vortices appear/disappear on a single plaquette, thereby interpolating between different topological sectors. The proliferation of such instantons at small $g$ has a momentous consequence. 

One is used to the fact that electromagnetic fields mediate Coulomb forces between static charges, a $1/|{\bf x}|$ potential in 3D, or a ${\rm log} |{\bf x}|$ potential in 2D. The monopole plasma in the compact version of QED, however, drastically changes this and generates a linearly confining potential $ \sigma |{\bf x}| $ between charges of opposite sign, where 
\begin{equation}
\sigma = {\hbar \omega_P \over \ell} \sqrt{16 \over \pi g \ell_0\omega_P }\ {\rm e}^{-{\pi g \over 2\ell_0 \omega_P}G}\ ,
\label{tension}
\end{equation}
is the string tension, $G=O(1)$, and we have reinstated physical units. This is the phenomenon of confinement, known from strong interactions (for a review, see \cite{greensite}), with electric fields playing the role of chromo-electric fields and Cooper pairs playing the role of quarks. An electric flux tube (string) dual to Abrikosov vortices holds together charges in neutral pion excitations. There is no charged excitation in the spectrum for arrays larger than the pion size, and this is the origin of the infinite electric resistance in this phase, dual to the infinite conductivity in the superconducting phase. This state of matter is known as a superinsulator, and was first predicted in \cite{dst} (for a review, see \cite{book}). Superinsulators \cite{dst, vinokur1, vinokur2, diamantini1, moncon}, with their divergent electrical resistance, have been experimentally detected in InO \cite{shahar}, TiN \cite{vinokur1}, NbTiN \cite{mironov}, and NbSi \cite{humbert}. A recent measurement of the dynamic response of superinsulators confirmed that the potential holding together $\pm$ charges is indeed linear \cite{dynamic}.

\section{Why Disorder and Dissipation Are Not Relevant for the Quantum Phases of Granular Films}

We are now in the position to explain why disorder is irrelevant for the quantum phases of granular superconductors. As mentioned in the introduction, JJAs are the closest ordered Hamiltonian system on which to add positional and size disorder of the granules. This disorder then affects the behaviour of the excitations in the spectrum of JJAs. In the topological insulator phase, disorder can indeed help to pin the excess bulk charges and vortices around the integer-filling ground state when $g$ deviates from its central value $g=1$. But even this is not necessary; these excitations form a Wigner crystal \cite{nodisorder}, exactly as can happen in the fractional quantum Hall effect \cite{kivelson}. And, indeed, the topological insulator state is clearly detected in perfectly ordered JJAs \cite{marcus}. Finally, the edge currents are symmetry-protected and thus transparent to Anderson localization. 

In the superinsulating phase, there are no U(1) charged excitations in the spectrum; they are confined (for large enough samples). As such, disorder may influence the neutral excitations responsible for the thermal properties, but not the electric transport properties. The infinite resistance (even at finite temperatures) is due exclusively to strong electric interactions first preventing Bose condensation and then becoming linearly confining by an instanton plasma. This is analogous to the situation in the disordered Schwinger model (1D QED), where confinement is kinematic. First, one has to identify the correct spectrum of neutral  mesons and only then can one even speak of disorder \cite{sondhi}. Of course, disorder can influence itself the strength of the Coulomb interaction, but this leads only to a renormalization of $E_C$, i.e., of $g$ \cite{finkelstein}. 

Finally, dissipation by single-electron tunnelling is often mentioned as a relevant phenomenon for granular superconductors. However, in the bosonic topological insulator phase, the only possible dissipation is due to quantum phase slips, since edge states are symmetry protected. In the superinsulating phase, there are no single electrons in the spectrum; they are confined too. The only region where single-electron dissipation may become relevant is at higher temperatures, near the thermal transition, where the string tension becomes very small, in perfect analogy to the dual superconductors. 

\section{Why the Full Phase Structure Is Not Yet Seen in JJAs}
The $g\ge1$ ($R \le R_Q$) segment of the bosonic topological insulator phase (failed superconductor) has been recently experimentally detected in Al/InAs JJAs, confirming that disorder is irrelevant for this phase \cite{marcus}. As mentioned in the introduction, this has been achieved by a voltage gate, which depletes the available charges for tunnelling, which amounts essentially to varying $E_J$ at fixed $E_C$. Apparently, it is not possible to reach low enough values of $g$ by this procedure alone, without destroying superconductivity of the Al islands themselves. In our opinion, one should consider a more regular arrangement of superconducting islands and exposed semiconductor substrate, forming something like a checkerboard with, say, the white squares being the superconducting islands and the black squares the ``dual'' islands of exposed substrate, and introduce a second, separate procedure to govern the dual, vortex tunnelling coupling $E_C$.  We predict that once this can be achieved, the described phase structure will be exposed also for completely ordered JJAs.

\end{document}